\documentclass[prl,twocolumn,showpacs,aps]{revtex4-1}
\usepackage{graphicx}
\usepackage{dcolumn}
\usepackage{bm}
\usepackage{amsmath,amssymb,amsfonts}
\usepackage{latexsym}
\usepackage{bbm}
\usepackage{color}
\usepackage{epsfig}

\begin{document}

\title{Magnetic mass effect on the sphaleron energy}

\author{Koichi Funakubo$^1$}
\email{funakubo@cc.saga-u.ac.jp}
\author{Eibun Senaha$^{2,3}$}
\email{eibun.senaha@tdtu.edu.vn}
\affiliation{$^1$Department of Physics, Saga University, Saga 840-8502 Japan}
\affiliation{$^2$Theoretical Particle Physics and Cosmology Research Group, Advanced Institute of Materials Science, Ton Duc Thang University, Ho Chi Minh City 700000, Vietnam}
\affiliation{$^3$Faculty of Applied Sciences, Ton Duc Thang University, Ho Chi Minh City 700000, Vietnam}
\bigskip

\date{\today}

\begin{abstract}
We elucidate a magnetic mass effect on a sphaleron energy that is crucial for baryon number preservation needed for successful electroweak baryogenesis. We find that the sphaleron energy increases in response to the magnetic mass. As an application, we study the sphaleron energy and electroweak phase transition with the magnetic mass in a two-Higgs-doublet model. Although the magnetic mass can screen the gauge boson loops, it relaxes a baryon number preservation criterion more effectively, broadening the baryogenesis-possible region. Our findings would be universal in any new physics models as long as the gauge sector is common to the standard model.
\end{abstract}


\maketitle

\paragraph{Introduction.---}
A sphaleron~\cite{Manton:1983nd,Klinkhamer:1984di} in electroweak (EW) theories plays a central role in the baryon and lepton number-violating process at high temperature and its rate in the EW-symmetry-broken phase 
is crucial to successful electroweak baryogenesis (EWBG)~\cite{ewbg}. In particular, energy of the sphaleron, which is predominantly scaled by a vacuum expectation value (VEV) of the Higgs field, has to be sufficiently large in order to preserve the baryon asymmetry, calling for a strong first-order electroweak phase transition (EWPT).

It is known that ordinary perturbative expansion at zero temperature would break down at high temperature due to uncontrollable temperature-dependent power corrections. The standard prescription for this problem is called thermal resummation; it reorganizes the perturbative expansion in a consistent manner. 
It is common practice in the literature that finite-temperature effective potential is improved by the resummation methods of Parwani~\cite{Parwani:1991gq} or Arnold-Espinosa~\cite{Arnold:1992rz}. 
With such a resummed effective potential, EWPT is studied in a plethora of models beyond the standard model (SM). Regarding the thermal resummation for the EW gauge sector, much attention has been given to electric masses, even though a magnetic mass of the $SU(2)$ gauge field can, in principle, enter the problems. It is thus legitimate to take both the resummation effects into consideration (for earlier studies of the SM, see Refs.~\cite{Espinosa:1992kf,Buchmuller:1993bq}).

In evaluating the sphaleron rate using a WKB method, it is necessary for consistency to use the same resummed Lagrangian that is applied to the EWPT study. Due to the fact that the sphaleron is a magnetic configuration, the magnetic mass term appears at the lowest order in the equation of motion for the sphaleron, while the electric masses come into the gauge boson loops in the resummed effective potential, which are higher order. Regardless of the potential significance of the magnetic mass, a devoted study has been missing so far. Since the required strength of the first-order EWPT is predominantly determined by the sphaleron energy, ramifications of the magnetic mass effect is of great importance for successful EWBG.

In this paper, we investigate the magnetic mass effect on the sphaleron energy 
and quantify its impact on the baryon number preservation criterion (BNPC) needed for EWBG.
For illustrative purposes, we first work on the SM at zero temperature
and clarify to what extent the sphaleron energy can be affected by the presence of the magnetic mass.
Given the fact that the magnetic mass is inherently nonperturbative and its robust estimate seems unavailable so far, we thus allow it to vary within a reasonable range studied in the literature. Then, as a realistic application, we consider an EWBG scenario in a general two-Higgs-doublet model (2HDM) (for a recent study, see, e.g., Ref.~\cite{rhottEWBG}) and evaluate the sphaleron energy and BNPC with the magnetic mass at a critical temperature of EWPT. Our analysis shows that the sphaleron energy is the increasing function against the magnetic mass,
which renders BNPC more relaxed and EWBG-possible regions get broadened accordingly. 
This conclusion would apply to any model beyond the SM as long as the gauge sector is common to the SM.

\paragraph{Gauge-invariant mass term.---}
We first delineate how the magnetic mass is implemented in our problem.
Here we demonstrate two different approaches and show that they both lead to the same mass form under a certain condition.

In general, the bilinear term of the gauge field takes the form
\begin{align}
\mathcal{L}_{\text{eff}}^{(2)} &=  \text{Tr}\big[A^\mu \Pi_{\mu\nu} A^\nu\big]
= \frac{1}{2}A^{a\mu}\Pi_{\mu\nu}A^{a\nu},\label{L2}
\end{align}
where $\Pi_{\mu\nu}$ is the self-energy and its general form in the momentum space is cast into the form~\cite{Buchmuller:1993bq}
\begin{align}
\Pi_{\mu\nu}(p) &= \Pi_L(p)L_{\mu\nu}(p)+ \Pi_T(p)T_{\mu\nu}(p)+\Pi_G(p)G_{\mu\nu}(p) \nonumber\\
&\quad+\Pi_S(p)S_{\mu\nu}(p),
\end{align}
where $L_{\mu\nu}(p) = u_\mu^Tu_\nu^T/(u^T)^2$, $T_{\mu\nu}(p) = g_{\mu\nu}-p_\mu p_\nu/p^2-L_{\mu\nu}(p)$, $G_{\mu\nu}(p)=p_\mu p_\nu/p^2$, $S_{\mu\nu}(p) = (p_\mu u_\nu^T+p_\nu u_\mu^T)/\sqrt{(p\cdot u)^2-p^2}$ with $u_\mu^T=u_\mu-p_\mu(u\cdot p)/p^2$ and $u_\mu$ specifies the thermal bath.
We consider a static case in which $p^0=0$ in the rest frame of the thermal bath [$u_\mu=(1,\boldsymbol{0})$]. In this case, under the condition of $\partial_iA_i=0$, which is satisfied by the sphaleron ansatz adopted here, the bilinear Lagrangian (\ref{L2}) is reduced to the local form
\begin{align}
\mathcal{L}_{\text{eff}}^{(2)} = \frac{1}{2}\Pi_L(A_0^a)^2-\frac{1}{2}\Pi_T(A_i^a)^2.\label{L2local}
\end{align}
Since the sphaleron is the magnetic object as mentioned above, only $\Pi_T$ is involved in the equations of motion for the sphaleron. It is well known that the computation of $\Pi_T$ requires a nonperturbative approach~\cite{Gross:1980br}. We do not attempt to do this here, and we treat it as an input parameter.
We note that the non-Abelian gauge invariance requires higher-order terms in $A_\mu$.
In the case of a hard thermal loop approximation in QCD, effective $n$-point ($n\ge 3$) functions of $A_\mu$ would be zero to the leading order, and only the $\Pi_L$ term in Eq.~(\ref{L2local}) survives in the static limit~\cite{Blaizot:2001nr}. 
In contrast, we are not aware of the counterpart in effective theories of the magnetic mass.

In contrast to this approach, the second one is manifestly gauge invariant by construction. 
As discussed in Ref.~\cite{Zwanziger:1990tn}, one of the gauge-invariant operators with a mass dimension of two 
can be constructed by minimizing the volume integral of $\text{Tr}[A_\mu^2]$ along gauge orbits, i.e.,
\begin{align}
\int d^4x~A_{\text{min}}^2=\mathop{\text{min}}_{\{U\}}\int d^4x~{\rm Tr}\big[(A_\mu^{U})^2\big],
\label{Amin2}
\end{align}
where $U$ is the gauge transformation function, that is, $A^U_\mu = UA_\mu U^{-1}+\frac{i}{g}(\partial_\mu U)U^{-1}$ with $g$ being a gauge coupling.
As proven, $A_{\text{min}}^2$ can be expressed by an infinite series of nonlocal and gauge-invariant terms
constructed by the covariant derivative and field strength of $A_\mu$ such as $\text{Tr}[F_{\mu\nu}(D^2)^{-1}F^{\mu\nu}]$~\cite{Zwanziger:1990tn} (for a dedicated study of Yang-Mill theories with this nonlocal term, see Ref.~\cite{Capri:2005dy}). It is shown that $A_{\text{min}}^2$ can be local if $\partial_\mu A^\mu=0$ is satisfied~\cite{Zwanziger:1990tn,Capri:2005dy}, which greatly simplifies our calculation as delineated below.

\paragraph{Sphaleron with magnetic mass.---}
Now we work out the sphaleron energy in the presence of the magnetic mass in the SM. As demonstrated in Ref.~\cite{sphwU1}, the $U(1)_Y$ contribution is rather minor, so we do not include it in the estimate of the sphaleron energy.

The original ansatz of the sphaleron adopted in Ref.~\cite{Klinkhamer:1984di} causes a divergence in the $A_i^2$ term in the energy functional of the sphaleron. To avoid this, we perform an $SU(2)$ gauge transformation as
$A_\mu \to  V A_\mu V^{-1} + \frac{i}{g}(\partial_\mu V)V^{-1}$ with $V$ being the inverse of $U^\infty$ (for the explicit form of $U^\infty$, see Ref.~\cite{Klinkhamer:1984di}). Furthermore, after a rigid $SU(2)$ transformation $U^\infty \to U_LU^\infty U_R$ (for $U_{L,R}$, see~Ref.~\cite{Klinkhamer:1984di}), the ansatz of $A_i$ is cast into the form
\begin{align}
A_i =  -\frac{1-f(r)}{gr}\epsilon_{ija}\hat{x}_j\tau^a,
\end{align} 
where $\epsilon_{ija}$ is the Levi-Civita symbol, $\hat{x}_j=x_j/r$ with $r=\sqrt{x_1^2+x_2^2+x_3^2}$,
$\tau^a$ are the Pauli matrices and $f(r)$ denotes the profile function. 
Unlike the original sphaleron ansatz, $A_i$ is proportional to $(1-f)$, which is damped exponentially at $r\to \infty$,
and thus the $A_i^2$ term becomes finite. 
Most importantly, the above ansatz satisfies $\partial_iA_i=0$. Exploiting this property as well as the $A_0=0$ gauge, which is often adopted for finding a static classical solution, Eq.~(\ref{Amin2}) is reduced to the local form; thereby the magnetic mass contribution to the energy function of the sphaleron takes the form
\begin{align}
\Delta E_{\text{sph}}= \frac{m_T^2}{2}\int d^3\boldsymbol{x}~A_i^aA_i^a.
\label{Esph_mT}
\end{align}
Adding this energy shift into the ordinary energy functional of the sphaleron in the SM~\cite{Klinkhamer:1984di}, 
one arrives at
\begin{align}
E_\text{sph}&=\frac{4\pi v}{g}\int_0^\infty d\xi~
\bigg[
	4f'^2+\frac{8}{\xi^2}(f-f^2)^2+\frac{\xi^2}{2}h'^2 \nonumber\\
&\hspace{2.5cm} 
	+(h^2+r_m^2)(1-f)^2+\frac{\xi^2V_0(h)}{g^2v^4}
\bigg],\label{Esph_func}
\end{align}
where $\xi=gvr$ with $v$ being the VEV of the Higgs field, 
$V_0(h)=\lambda v^4(h^2-1)^2/4$, and $r_m=m_T^{}/(gv/2)=m_T^{}/m_W^{}$.
From the above energy functional, it follows that
\begin{align}
\frac{d^2 f}{d\xi^2} & = \frac{2}{\xi^2}(f-f^2)(1-2f)-\frac{1}{4}(h^2+r_m^2)(1-f), 
\label{EOM_f}\\
\frac{d^2h}{d\xi^2}
&= -\frac{2}{\xi}\frac{dh}{d\xi}+\frac{2}{\xi^2}h(1-f)^2+\frac{1}{g^2v^4}\frac{\partial V_0}{\partial h}, \label{EOM_h}\end{align}
and the boundary conditions are $\lim_{\xi\to0} f(\xi) = 0$, $\lim_{\xi\to0} h(\xi) = 0$, $\lim_{\xi\to\infty} f(\xi) = 1$, and $\lim_{\xi\to\infty} h(\xi) = 1$.
Since the magnetic mass correction is positive definite, we expect that the resulting sphaleron would be more massive than the usual one. Moreover, $r_m^2$ could exceed unity unlike the $h^2$ term.

\begin{figure}[t]
\center
\includegraphics[width=7.3cm]{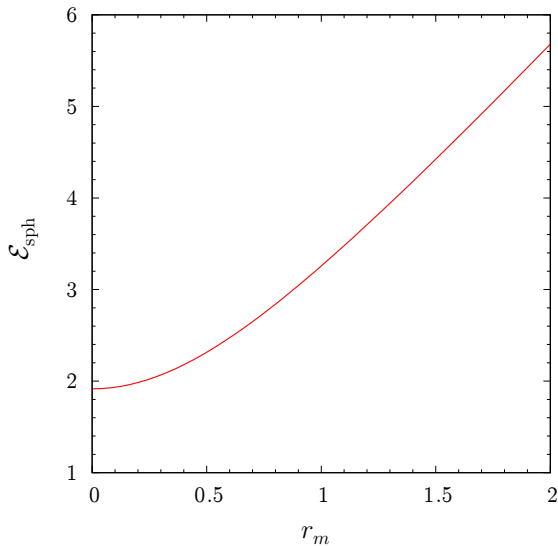}
\caption{Dimensionless sphaleron energy $\mathcal{E}_{\text{sph}}=E_{\text{sph}}(g/4\pi v)$ as a function of $r_m=m_T^{}/m_W^{}$ in the SM without the $U(1)_Y$ contribution. Here, $\mathcal{E}_{\text{sph}}=1.92$ for $r_m=0$ agrees with the result of Klinkhamer and Manton~\cite{Klinkhamer:1984di}.}
\label{fig:Esph_rm}
\end{figure}
In Fig.~\ref{fig:Esph_rm}, a dimensionless sphaleron energy defined by $\mathcal{E}_{\text{sph}}=E_{\text{sph}}(g/4\pi v)$ is plotted as a function of $r_m$. We use the values of $m_W=80.4$ GeV, $m_h=125$ GeV, and $v=246$ GeV to fix the input parameters.
One can see that $\mathcal{E}_{\text{sph}}$ increases from 1.92 to 5.68 as $r_m$ varies from 0 to 2. This drastic change stands in stark contrast to the dependence of the Higgs mass on $\mathcal{E}_{\text{sph}}$. As first clarified in Ref.~\cite{Klinkhamer:1984di}, $\mathcal{E}_{\text{sph}}$ is an increasing function of the Higgs mass, which would be at most around 2.7 even when the Higgs mass goes to infinity.
The question to be answered is how large $r_m$ could be in a realistic case.
As widely studied, the magnetic mass is given by $m_T^{}=cg^2T$, with $c$ being a coefficient that is determined by nonperturbative methods. For instance, in early studies~\cite{Espinosa:1992kf,Buchmuller:1993bq} it is found in the $SU(2)$ gauge-Higgs model that $c = 1/(3\pi)\simeq 0.11$ by solving a gap equation in the high-$T$ limit. However, this result turns out to be gauge dependent. Estimates using the gauge-invariant gap equations provide somewhat bigger values: $c=0.28$~\cite{Buchmuller:1994qy}, $c=0.38$~\cite{Alexanian:1995rp}, and $c=0.35$~\cite{Patkos:1997cw}. 
On the other hand, a lattice calculation finds an even bigger value $c=0.46$~\cite{Heller:1997nqa} with a caveat~\cite{Cucchieri:2001tw}. Given the lack of a convincing estimate, we treat $c$ as the varying parameter and take the maximal value of $c=0.45$ in our analysis.
It should be noted that the degree to which $\mathcal{E}_{\text{sph}}$ is enhanced depends not only on the value of $m_T^{}$ but also on that of $m_W(T)=gv(T)/2$ at finite temperature $T$.
Generally, $r_m$ gets bigger for weaker first-order EWPT because of  $r_m=2c g(T_C/v_C)$.
For $c=0.45$ and $v_C/T_C=0.3$, one would get $r_m \simeq 2$, where $g=0.65$ is used.  
Therefore, $\mathcal{E}_{\text{sph}}$ could be 3 times larger than the case without the magnetic mass effect.

Now we discuss implications of the enhanced $\mathcal{E}_{\text{sph}}$ for EWBG.
As alluded to above, the baryon and lepton number-changing process has to shut off in the EW broken phase, which yields BNPC as~\cite{Bochkarev:1987wf,Arnold:1987mh,Funakubo:2009eg}
\begin{align}
\frac{v}{T}>\frac{g}{4\pi\mathcal{E}_{\text{sph}}}\Big[44.35+\mbox{log corrections}\Big] \equiv \zeta_{\text{sph}}.\label{BNPC}
\end{align}
where the logarithmic corrections come from the fluctuation about the sphaleron such as zero mode factors, which could be 10\% correction in the minimal supersymmetric SM~\cite{Funakubo:2009eg}. 
Since $\zeta_{\text{sph}}$ is predominantly affected by $\mathcal{E}_{\text{sph}}$, we do not consider the subdominant logarithmic corrections hereafter.
Without large supercooling, $T$ can be set to a critical temperature $T_C$ at which the Higgs potential has two degenerate minima. As seen from Eq.~(\ref{BNPC}), the larger $\mathcal{E}_{\text{sph}}$ lowers $\zeta_{\text{sph}}$. One can find $\zeta_{\text{sph}}=1.20-0.41$ for $\mathcal{E}_{\text{sph}}$ obtained in Fig.~\ref{fig:Esph_rm}. Therefore, the commonly used BNPC in the literature, $v_C/T_C>1$,
does not always give a reasonable criterion in the presence of the magnetic mass.
One remark here is that as emphasized in Ref.~\cite{Funakubo:2009eg} (see also Ref.~\cite{Braibant:1993is} for an early study), $\mathcal{E}_{\text{sph}}(T=T_C)$ tends to be smaller than $\mathcal{E}_{\text{sph}}(T=0)$, implying that the above values of $\zeta_{\text{sph}}$ could be underestimated to some extent. 
We therefore evaluate $\mathcal{E}_{\text{sph}}(T_C)$ explicitly in the case of 2HDM, which is one of the simplest extended models that accommodates the EWBG possibility.

\paragraph{Application to 2HDM.---}
Here, we consider a general 2HDM in which an \textit{ad hoc} $Z_2$ symmetry is not imposed. For the sake of simplicity, we take an alignment limit in which $\sin(\beta-\alpha)=1$, where $\alpha$ and $\beta$ are the mixing angles between two neutral Higgs bosons and two Higgs VEVs, respectively. In this case, the Higgs couplings to gauge bosons and fermions are reduced to the SM values at tree level. In addition, we put another simplifying assumption that $\tan\beta=1$ and $Z_2$-breaking Higgs quartic couplings are taken to zero to a first approximation, while flavor-changing neutral Higgs couplings in the Yukawa sector are allowed to exist in order to provide $CP$ violation for baryogenesis~\cite{rhottEWBG}. In this setup, the effective potential at tree level is SM like. 
We emphasize here that the above simplification does not spoil the core of our discussion, and the main findings would not be affected by it.

Following the resummation method of Parwani~\cite{Parwani:1991gq}, we construct the resummed Lagrangian as $\mathcal{L}=(\mathcal{L}_R-\Delta m^2(T)\Psi^2)+(\mathcal{L}_{\text{CT}}+\Delta m^2(T)\Psi^2)$, where $\mathcal{L}_R$ and $\mathcal{L}_{\text{CT}}$ denote the renormalized Lagrangian and corresponding counterterms, respectively. The thermal mass terms of the particles in question are collectively denoted as $m^2(T)\Psi^2$ [for explicit forms of $m^2(T)$, see, e.g., Ref.~\cite{Cline:1996mga}]. 
The thermal masses in the first parentheses are treated as zeroth order in the resummed perturbation theory, while those in the second one are treated as the part of the counterterms.
For the magnetic mass, we expand the gauge-invariant operator $\int d^4x A_{\text{min}}^2$ as
\begin{align}
m_T^2\int d^4x~A_{\text{min}}^2\simeq \frac{m_T^2}{2}\int d^4x~A^{a\mu} T_{\mu\nu}A^{a\nu}+\cdots,
\end{align}
where the ellipsis denotes higher-order terms in $A_\mu$. We retain only the quadratic term that is needed for the one-loop effective potential calculation.
Taking the Landau gauge and regularizing the resummed effective potential using the $\overline{\text{MS}}$ scheme, one finds
\begin{align}
V_1(\varphi; T) &= \sum_i n_i
\left[
	\frac{\bar{m}^4_i}{64\pi^2}\left(\ln\frac{\bar{m}^2_i}{\bar{\mu}^2}-c_i\right)
	+\frac{T^4}{2\pi^2}I_{B,F}(a_i^2)
\right],
\label{V1}
\end{align}
where $\bar{m}_i$ denote the thermally corrected $\varphi$-dependent masses of the species of $i$ and $a_i^2 = \bar{m}^2_i/T^2$, with $i=h,H,A,H^{\pm}$ (Higgs bosons), $G^0, G^\pm,$ [Nambu-Goldstone (NG) bosons)]  $W_{L,T}^{\pm}, Z_{L,T}^{}, \gamma_{L,T}^{}$ (longitudinal and transverse gauge bosons), $t$ (top quark), and $b$ (bottom quark), respectively. Note that $\bar{\mu}$ is a scale determined by renormalization conditions described below; $n_i$ are the degrees of freedom, and their signs are determined by the statistics of the particles; and $c_i$ are 3/2 for the scalars, longitudinal gauge bosons, and fermions, while 1/2 for the transverse gauge bosons. Here, $I_{B,F}(a^2)$ are the one-loop thermal functions for bosons and fermions, which are, respectively, given by $I_{B,F}(a^2)=\int_0^\infty dx~x^2 \ln\Big[1\mp \exp[-\sqrt{x^2+a^2}]\Big]$. Their numerical evaluations are made using fitting functions described in Ref.~\cite{Funakubo:2009eg}.

For the renormalization of the vacuum and the mass of the SM-like Higgs ($h$) at one-loop level, we employ a scheme in which the tree-level relations are not altered by the one-loop contributions~\cite{Kirzhnits:1976ts}. A thorny problem in this scheme is that the massless NG bosons cause an infrared divergence that calls for resummation of higher-order corrections~\cite{NGresum}. It is demonstrated in Ref.~\cite{Chiang:2018gsn} that such a NG resummation has little effect on EWPT, so we do not take the NG bosons into account in our numerical study. Moreover, we do not include the SM-like Higgs boson loop contribution either since its treatment is somewhat technical in the case of $\bar{m}_h^2<0$, even though its numerical impact is rather minor. 
With that in mind, we calculate $v_C/T_C\equiv \mathcal{R}_C$ using Eq.~(\ref{V1}) plus the tree-level Higgs potential. Likewise, $\mathcal{E}_{\text{sph}}$ is evaluated by using the same resummed effective potential at $T_C$, replacing $V_0$ in Eqs.~(\ref{Esph_func}) and (\ref{EOM_h}) with $V_0+V_1$. 

\begin{figure}[t]
\center
\includegraphics[width=7cm]{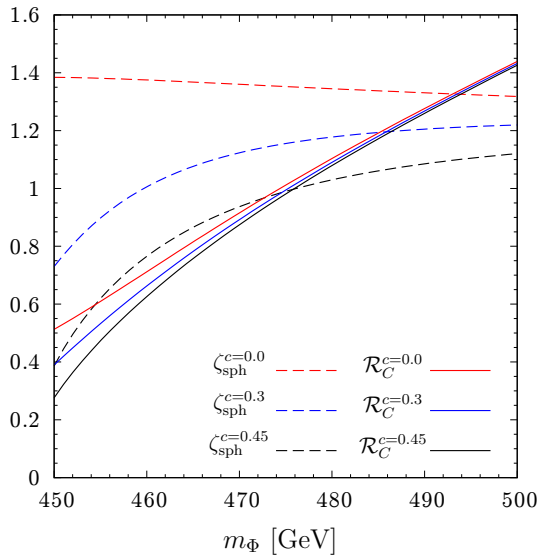}
\caption{$\mathcal{R}_C=v_C/T_C$ and $\zeta_{\text{sph}}$ as functions of $m_\Phi$ with $c=0.0$, 0.3, and 0.45, where $m_\Phi=m_H=m_A=m_{H^\pm}$. We take  $\sin(\beta-\alpha)=\tan\beta=1$ and $M=300$ GeV. The magnetic mass is parametrized as $m_T^{}=cg^2T$.}
\label{fig:EWPT_mPhi}
\end{figure}
For illustration, we consider a case in which all the heavy Higgs boson masses are degenerate in masses, $m_\Phi\equiv m_H=m_A=m_{H^\pm}$, and $M^3=m_3^2/(\sin\beta\cos\beta)=(300~\text{GeV})^2$, with $m_3^2$ being the squared mixing mass between the two Higgs doublets in a generic basis where both Higgs doublets develop the VEVs.

In Fig.~\ref{fig:EWPT_mPhi}, $\mathcal{R}_C\equiv v_C/T_C$ and $\zeta_{\text{sph}}$ are shown as functions of $m_\Phi$ varying from 450 GeV to 500 GeV. The solid curves represent $\mathcal{R}_C$ in the cases of $c=0.0$ (red), 0.3 (blue), and 0.45 (black).  In any case, $\mathcal{R}_C$ gets enhanced when $m_\Phi$ is considerably bigger than $M$, which is due to the fact that thermal loops of the heavy Higgs bosons enhance the potential barrier. The differences among the three cases become more pronounced in the lower $\mathcal{R}_C$ region, where the gauge boson thermal loops are the main contributors to the potential barrier and thus the screening effect by the magnetic mass is more influential. The three dashed curves display $\zeta_{\text{sph}}$ for $c=0.0$, 0.3, 0.45 with the same color scheme as $\mathcal{R}_C$. Without the magnetic mass effect, it is found that $\zeta_{\text{sph}}=1.38-1.32$ in which $\mathcal{E}_{\text{sph}}(T_C)=1.67-1.75$, which is smaller than $\mathcal{E}_{\text{sph}}=1.92$ found in the case of the SM at $T=0$ discussed above. The nonzero magnetic mass cases show more dramatic effects on $\zeta_{\text{sph}}$ compared to $\mathcal{R}_C$, which is attributed to the substantial enhancement of $\mathcal{E}_{\text{sph}}$ as expected from the analysis in the SM. As is the case of $\mathcal{R}_C$, the magnetic mass effect on $\zeta_{\text{sph}}$ gets more pronounced in the lower $\mathcal{R}_C$ region and its change is even more drastic, which is due to the enhancement of $r_m=2cg/(\pi\mathcal{R}_C)$. For $c=0.45$ and $m_\Phi=450$ GeV, it is found that $r_m = 2.1$. From the EWBG point of view, the region of $\mathcal{R}_C>\zeta_{\text{sph}}(T_C)$ is relevant, which is satisfied if $m_\Phi \gtrsim 493$ GeV for $c=0.0$, $m_\Phi \gtrsim 486$ GeV for $c=0.3$, and $m_\Phi \gtrsim 476$ GeV for $c=0.45$, expanding the domain of the EWBG-possible regions.

We make a couple of remarks on our results.
First, although the magnetic mass effect becomes sizable for $\mathcal{R}_C\ll1$, the perturbative treatment of EWPT starts to lose its reliability, calling for nonperturbative approaches, while our primary interest is EWBG in which $\mathcal{R}_C> \zeta_{\text{sph}}(T_C)$. Second, in the current approach, the unambiguous numerical estimate is plagued by uncertainties of the magnetic mass. Nevertheless, the estimated values of $c$ in the literature consistently lie in the rather restricted range $c\simeq 0.3-0.45$~\cite{c_fermion}, and its impact on the sphaleron energy is already pronounced even for the conservative choice, $c=0.3$, as shown in Fig.~\ref{fig:EWPT_mPhi}. Besides, it could be possible for the magnetic mass to receive VEV-dependent corrections in the broken phase, as discussed in Ref.~\cite{Espinosa:1992kf}. We find that the correction discussed there is positive and increases the sphaleron energy even further, though the robust numerical estimate cannot be made until a refined calculation is available. All in all, it is surely necessary to exercise caution when interpreting our results, but even after taking such uncertainties into consideration, the overlooked effect is robustly present and influential in existing studies. Lastly, it is known in the SM that the sphaleron rate estimated by the lattice calculation is smaller than that by the perturbative method~\cite{Moore:1998swa}. By taking account of the magnetic mass effect in the latter, the discrepancy between them could be mitigated, which supports our results.

\paragraph{Conclusion.---}
Our analysis has unveiled that the sphaleron energy can increase considerably for the nonzero magnetic mass, which relaxes BNPC and broadens the parameter space for successful EWBG. The findings here would apply for any other new physics models as long as the gauge sector is common to the SM. 



\begin{thebibliography}{99}
\bibitem{c_fermion}
Those values are estimated in $SU(2)$ gauge theories without fermions. 
Given that the magnetic mass arises from the infrared region of the bosonic sector, it is unlikely that the current estimates of $c$ are drastically changed by the fermions as they act as massive modes.
\end{thebibliography}
\end{document}